\documentclass[iop,apj]{emulateapj}

\submitted{Received 2016 February 26; accepted 2016 April 20}
\journalinfo{The Astrophysical Journal}

\usepackage{amsmath,amssymb,amstext}

\usepackage[breaklinks,colorlinks,citecolor=black,linkcolor=black]{hyperref} 

\usepackage[all]{hypcap} 

\usepackage{aas_macros}
\usepackage{natbib}
\newcommand{\Ni}{\ensuremath{^{56}\mathrm{Ni}}}
\newcommand{\Co}{\ensuremath{^{56}\mathrm{Co}}}
\newcommand{\Fe}{\ensuremath{^{56}\mathrm{Fe}}}
\newcommand{\Mej}{\ensuremath{M_\mathrm{ej}}}
\newcommand{\Eej}{\ensuremath{E_\mathrm{ej}}}
\newcommand{\Msun}{\ensuremath{M_\odot}}
\newcommand{\vw}{\ensuremath{v_w}}
\newcommand{\Mdot}{\ensuremath{\dot{M}}}
\newcommand{\kmps}{\ensuremath{\mathrm{km~s^{-1}}}}

\shorttitle{Type~I\MakeLowercase{bn} supernovae}
\shortauthors{Moriya \& Maeda}

\begin{document}

\title{Circumstellar and explosion properties of Type~I\MakeLowercase{bn} supernovae}
\author{Takashi J. Moriya$^{1,2}$ and Keiichi Maeda$^{3,4}$}
\altaffiltext{1}{Division of Theoretical Astronomy, National Astronomical Observatory of Japan, 2-21-1 Osawa, Mitaka, Tokyo 181-8588, Japan; takashi.moriya@nao.ac.jp}
\altaffiltext{2}{Argelander Institute for Astronomy, University of Bonn, Auf dem H\"ugel 71, 53121 Bonn, Germany}
\altaffiltext{3}{Department of Astronomy, Kyoto University, Kitashirakawa-Oiwake-cho, Sakyo-ku, Kyoto 606-8502, Japan}
\altaffiltext{4}{Kavli Institute for the Physics and Mathematics of the Universe (WPI), The University of Tokyo Institutes for Advanced Study, The University of Tokyo, Kashiwanoha 5-1-5, Kashiwa, Chaba 277-8583, Japan}

\begin{abstract}
We investigate circumstellar and explosion properties of Type~Ibn supernovae (SNe) by analyzing their bolometric light curves. Bolometric light curves of Type~Ibn SNe generally have a large contrast between peak luminosity and late-phase luminosity, which is much larger than those of $^{56}$Ni-powered SNe. Thus, most of them are likely powered by the interaction between SN ejecta and dense circumstellar media. In addition, Type~Ibn SNe decline much faster than Type~IIn SNe, and this indicates that the interaction in Type~Ibn SNe ceases earlier than in Type~IIn SNe. Thus, we argue that Type~Ibn SN progenitors experience high mass-loss rates in a short period just before explosion, while Type~IIn SN progenitors have high mass-loss rates sustained for a long time. Furthermore, we show that rise time and peak luminosity of Type~Ibn and Type~IIn SNe are similar and thus, they have similar explosion properties and circumstellar density. The similar circumstellar density in the two kinds of SNe may indicate that mass-loss rates of Type~Ibn SN progenitors are generally higher than those of Type~IIn as the wind velocities inferred from narrow spectral components are generally higher in Type~Ibn SNe. We also show that \Ni\ mass and explosion energy of Type~Ibn SNe may be smaller than those of other stripped-envelope SNe, probably because they tend to suffer large fallback or some of them may not even be terminal stellar explosions.
\end{abstract}

\maketitle

\section{Introduction}
Some supernova (SN) progenitors are known to have extremely high mass-loss rates exceeding $\sim 10^{-4} \Msun~\mathrm{yr^{-1}}$ \citep[e.g.,][]{fox2011,kiewe2012,taddia2013,moriya2014}. Majority of them belongs to a class of Type~IIn SNe which have narrow hydrogen lines indicating the existence of dense circumstellar media (CSM) created by the progenitors' extensive mass loss. Their progenitors are also found to show variability in luminosity in decades to days before their explosions which is likely related to the formation of dense CSM \citep[e.g.,][]{ofek2013,ofek2014b,fraser2013}. Some SNe~IIn may not even be the terminal explosions of massive stars \citep[e.g.,][]{smith2011,kochanek2012}.

There also exist SNe showing strong helium narrow lines and they are classified as Type~Ibn SNe \citep{matheson2000,pastorello2007,pastorello2008,pastorello2008b,pastorello2015,pastorello2015b,pastorello2015e,pastorello2015f,pastorello2015c,pastorello2015d,foley2007,tominaga2008,mattila2008,smith2008,smith2012,immler2008,dicarlo2008,nozawa2008,anupama2009,sakon2009,sanders2013,gorbikov2014,turatto2014,modjaz2014,bianco2014}. SN~2006jc is one of the clear examples of this kind, with the extensively observed data set available for the first time \citep[e.g.,][]{pastorello2007,foley2007,tominaga2008,smith2008,immler2008,dicarlo2008,anupama2009,sakon2009}. The progenitor is found to have had a large luminosity increase two years before the explosion which is likely related to the formation of dense helium-rich CSM \citep{pastorello2007}. Type~Ibn SNe often show hydrogen emission which is weaker than their helium emission \citep[e.g.,][]{pastorello2008b,smith2012}. Combined with their pre-explosion variability, their progenitors are suggested to be massive stars in transition from luminous blue variables (LBVs) to Wolf-Rayet stars \citep{tominaga2008,smith2012,pastorello2015}. However, there exist diversities in Type~Ibn SNe and progenitors of Type~Ibn SNe may not be unique \citep[e.g.,][]{pastorello2015d}.

Thanks to recent transient surveys, the observed number of SNe~Ibn has increased dramatically in last years. However, because each SN~Ibn has their own characteristics, SNe~Ibn have been studied individually and there have not been systematic studies on their nature. In this paper, we collect bolometric light curves (LCs) of SNe~Ibn from literature and investigate general properties of SNe~Ibn to obtain their overall picture. We compare estimated properties of SN~Ibn progenitors to those of SNe~IIn and discuss the difference between them.

The rest of this paper is organized as follows. We first focus on the post-peak LC properties of SNe~Ibn in Section~\ref{sec:postpeak}. Then, we discuss the rise time and peak luminosity of SNe~Ibn in Section~\ref{sec:prepeak}. We summarize and conclude this paper in Section~\ref{sec:summary}.

\begin{figure*}
 \begin{center}
  \includegraphics[width=\columnwidth]{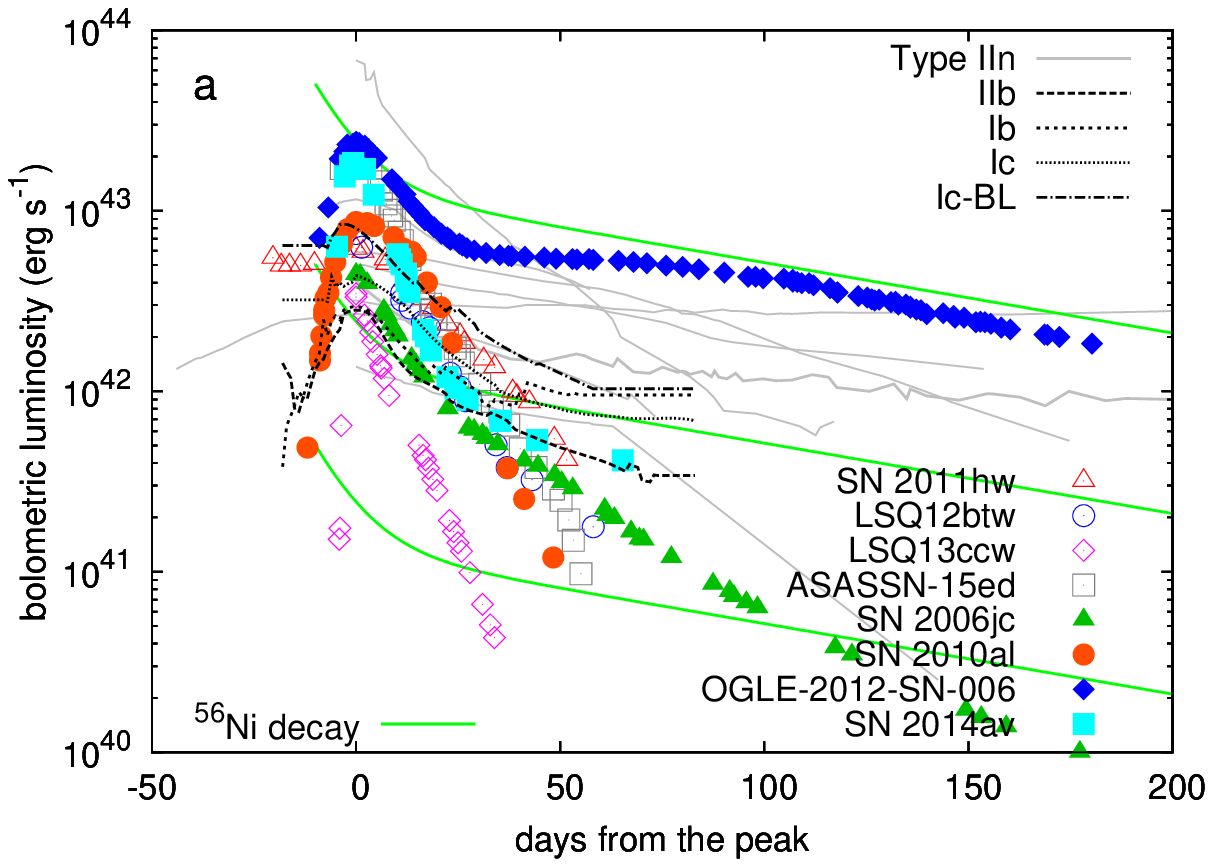}  
  \includegraphics[width=\columnwidth]{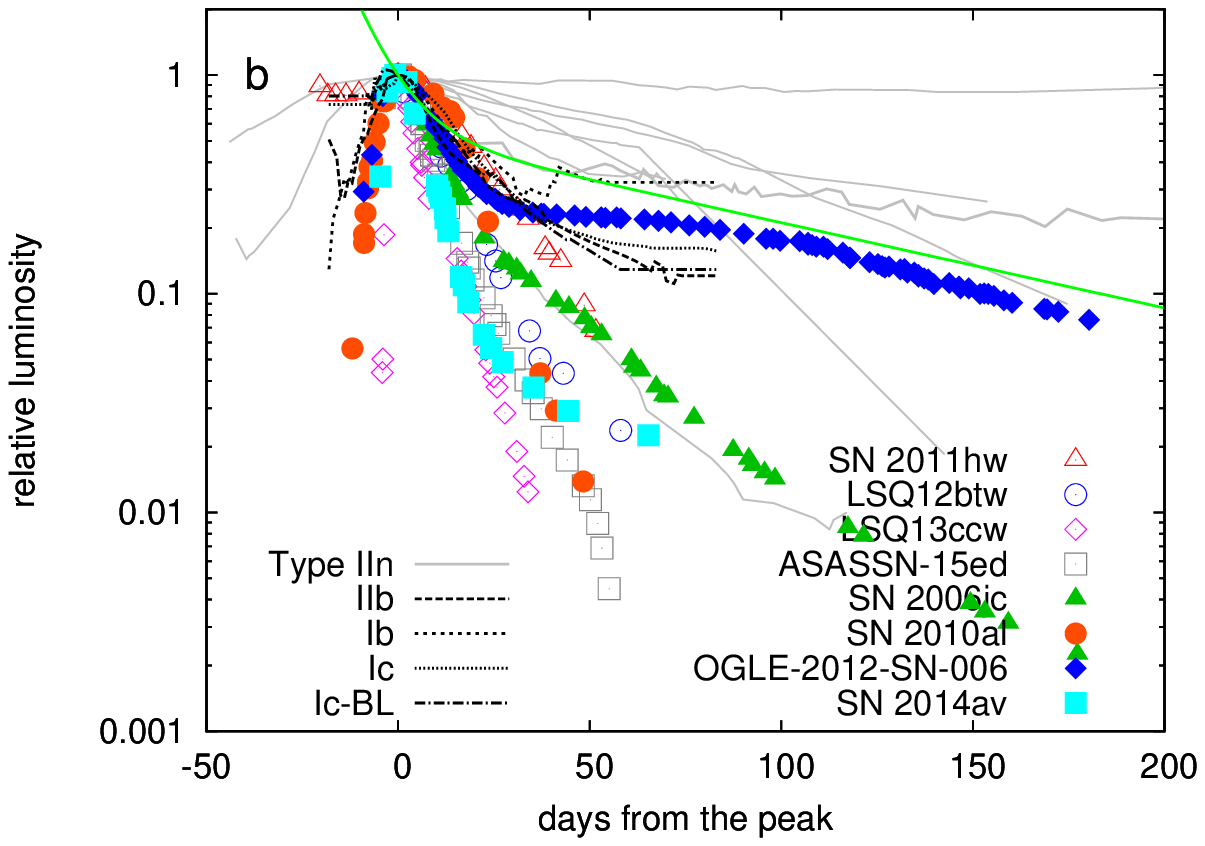}  
 \end{center}
\caption{
Bolometric LCs of SNe~Ibn, SNe~IIn, and stripped-envelope SNe. Bolometric LCs with filled symbols are constructed with infrared observations, while the others are not. Left panel shows the LCs as they are, and the right panel shows the same LCs scaled at their peak luminosity. The green solid lines are the total energy input from the nuclear decay of $\Ni\rightarrow\Co\rightarrow\Fe$. In the left panel, the green lines correspond to the nuclear decay of 1~\Msun\ (top), 0.1~\Msun\ (middle), and 0.01~\Msun\ (bottom) of \Ni. The time of explosion is assumed to be $-15$~days in the nuclear decay lines.
}
\label{fig:lightcurve}
\end{figure*}

\section{Post-peak light-curve properties}\label{sec:postpeak}
\subsection{Light-curve sample}\label{sec:samples}
We collect bolometric LCs of following SNe~Ibn from literature: SN~2006jc \citep{pastorello2007}, SN~2010al \citep{pastorello2015}, SN~2011hw \citep{pastorello2015}, LSQ12btw \citep{pastorello2015b}, OGLE-2012-SN-006 \citep{pastorello2015c}, LSQ13ccw \citep{pastorello2015b}, SN~2014av \citep{pastorello2015d}, and ASASSN-15ed \citep{pastorello2015e}. Figure~\ref{fig:lightcurve} shows the collected bolometric LCs.

Some SNe~Ibn are known to produce a large amount of dust and it is often essential to take the infrared luminosity contribution into account when their bolometric LCs are constructed \citep[e.g.,][]{tominaga2008}. The bolometric LCs of SN~2006jc, SN~2010al, OGLE-2012-SN-006, and SN~2014av are constructed by taking infrared contribution into account. These bolometric LCs are plotted with filled symbols in Fig.~\ref{fig:lightcurve}. The bolometric LCs of SN~2011hw, LSQ12btw, LSQ13ccw, and ASASSN-15ed are constructed without taking the infrared contribution into account, and they are shown with open circles in Fig.~\ref{fig:lightcurve}. Because the dust contribution becomes dominant from about 50~days after the LC peak \citep[e.g.,][]{pastorello2007,tominaga2008,nozawa2008,mattila2008}, their bolometric luminosity may be underestimated from around 50~days after the LC peak.

We collect bolometric LCs of the following SNe~IIn in \citet{taddia2013,stritzinger2012} for comparison: SNe~ 2005ip, 2005kj, 2006aa, 2006bo, 2006jd, 2006qq, and 2008fq. Additionally, we show bolometric LCs of SN~1998S \citep{fassia2000}, SN~2010jl (\citealt{zhang2012}, see also \citealt{fransson2014}), and SN~2011ht \citep{roming2012}. These bolometric LCs are studied in \citet{moriya2014} and their progenitor mass-loss histories are estimated there.

We also compare SN~Ibn LCs with stripped-envelope SN LCs. We take the template bolometric LCs of stripped-envelope SNe recently constructed by \citet{lyman2014}.

\subsection{What powers Type~Ibn SNe?}\label{sec:whatpowers}
Figure~\ref{fig:lightcurve}a shows the bolometric LCs collected in the previous section. We also show the available energy input from the nuclear decay of $\Ni\rightarrow\Co\rightarrow\Fe$ from initial \Ni\ masses of 1~\Msun\ (top), 0.1~\Msun\ (middle), and 0.01~\Msun\ (bottom) with green solid lines. The zero-point time of the nuclear decay, i.e., the explosion date, is set at $-15$~days in the figure. By comparing the peak luminosity and the nuclear decay lines, the amount of \Ni\ required to explain the peak luminosity of SNe~Ibn is found to be mostly between 0.1~\Msun\ and 1~\Msun, according to a simple estimate based on \citet{arnett1982}. However, the tails of the bolometric LCs indicate \Ni\ masses of below 0.1~\Msun\ in SNe~Ibn, which is inconsistent with those estimated from the peak luminosity.

Figure~\ref{fig:lightcurve}b shows bolometric LCs scaled at their peak luminosity. We first focus on the post-peak LC properties of SNe~Ibn and stripped-envelope SNe. Stripped-envelope SNe are powered by the radioactive decay of $\Ni\rightarrow\Co\rightarrow\Fe$. Their bolometric LCs constantly decline for about 30~days after the LC peak. Then, the LCs start to decline slower, roughly following the nuclear decay rate. This change of the LC slope to a slower decline is a general behavior of \Ni-powered SN LCs \citep[e.g.,][]{maeda2003}. However, most SNe~Ibn do not change their LC decline rates in the same timescale as stripped-envelope SNe do. SN~Ibn LCs have similar post-peak LC decline rates to those of stripped-envelope SNe for about 30~days after the LC peak. However, most of SNe~Ibn continue to decline without significant changes in decline rates. They keep declining and some of them eventually start to have slower decline at later time than stripped-envelope SNe (Fig.~\ref{fig:lightcurve}b). 

Figure~\ref{fig:lightcurve}b also clearly presents the existence of the large contrast between the peak luminosity and the late-phase luminosity in most SNe~Ibn. In stripped-envelope SNe, the contrast between the peak luminosity and the luminosity at about 50~days after the LC peak remains within a factor of 10 as is expected in \Ni-powered SNe. However, most SNe~Ibn have a much larger luminosity contrast.

Even if SNe are powered only by \Ni, there are several possible mechanisms to make a large luminosity contrast between the peak luminosity and late-phase luminosity. First, significant dust formation is observed in some SNe~Ibn. If the infrared contribution to the luminosity is not properly taken into account, the constructed luminosity will have an artificial luminosity decrease. The bolometric LCs of SNe~Ibn with filled symbols in Fig.~\ref{fig:lightcurve} take the infrared luminosity contribution into account, while the others do not. Even if we limit our samples to those with infrared, the above arguments on the luminosity contrast still holds. In addition, dust formation typically starts at around 50~days after the LC peak, but the rapid luminosity decrease is already found before around 50~days after the LC peak. Thus, we argue that the large luminosity contrast is not always caused by the missed infrared contribution, and we generally see it in SNe~Ibn. Another possible cause of the rapid decline is the insufficient gamma-ray trapping in SN ejecta. We discuss this possibility in Section~\ref{sec:expprop} and show that the large luminosity contrast is not due to the insufficient gamma-ray trapping.

To summarize the above arguments, the major heating source which powers the peak luminosity of SNe~Ibn is not \Ni. Especially, a large luminosity difference between the peak luminosity and the tail luminosity, which is not found in \Ni-powered stripped-envelope SNe, indicates that the radioactive decay of \Ni\ provides a little contribution to their bolometric LCs at least at their brightest phases. A rational guess to an alternative heating source is the interaction between SN ejecta and dense CSM as their spectral type and color indicate \citep[e.g.,][]{chugai2009}, although there can be exceptions like OGLE-2012-SN-006. We argue that the main heating source powering the peak luminosity of SNe~Ibn is the interaction in many cases.

\subsection{Interaction properties}
Even if SNe~IIn and SNe~Ibn are both powered by the interaction, the LC decline rates of SNe~Ibn are generally much faster than those of SNe~IIn (Fig.~\ref{fig:lightcurve}b). The slow LC declines of SNe~IIn are mainly due to ongoing interaction between dense CSM and SN ejecta. When SNe are powered by the continuing interaction, the bolometric LCs show a power-law decline \citep[e.g.,][]{chugai1994,moriya2013}. The slowly-declining LCs of SNe~IIn indicate that the interaction between SN ejecta and dense CSM continues more than 100~days after the LC peak, and their progenitors should have sustained large mass-loss rates for more than about 50~years before their explosions to make the extended distribution of the dense CSM \citep{moriya2014}.

The rapid LC declines in SNe~Ibn imply that their interaction between dense CSM and SN ejecta ceases quickly, in contrast to SNe~IIn. This LC rapid decline is not just due to the difference in the CSM compositions, as the interaction is a kinetic process and high-energy photons produced at the shock can be absorbed by heavier elements and re-emitted. Indeed, the luminosity by the interaction stays high as long as the interaction continues in numerical simulations of the interaction between SN ejecta and hydrogen-free CSM \citep[e.g.,][]{chugai2009,sorokina2015}. Thus, we argue that the dense CSM in SNe~Ibn are not extended as much as in SNe~IIn and only exist near their progenitors. This means that SN~Ibn progenitors are likely to have large mass-loss rates only shortly before their explosion, contrary to long-sustained large mass-loss rates in SN~IIn progenitors.

\begin{figure*}
 \begin{center}
  \includegraphics[width=\columnwidth]{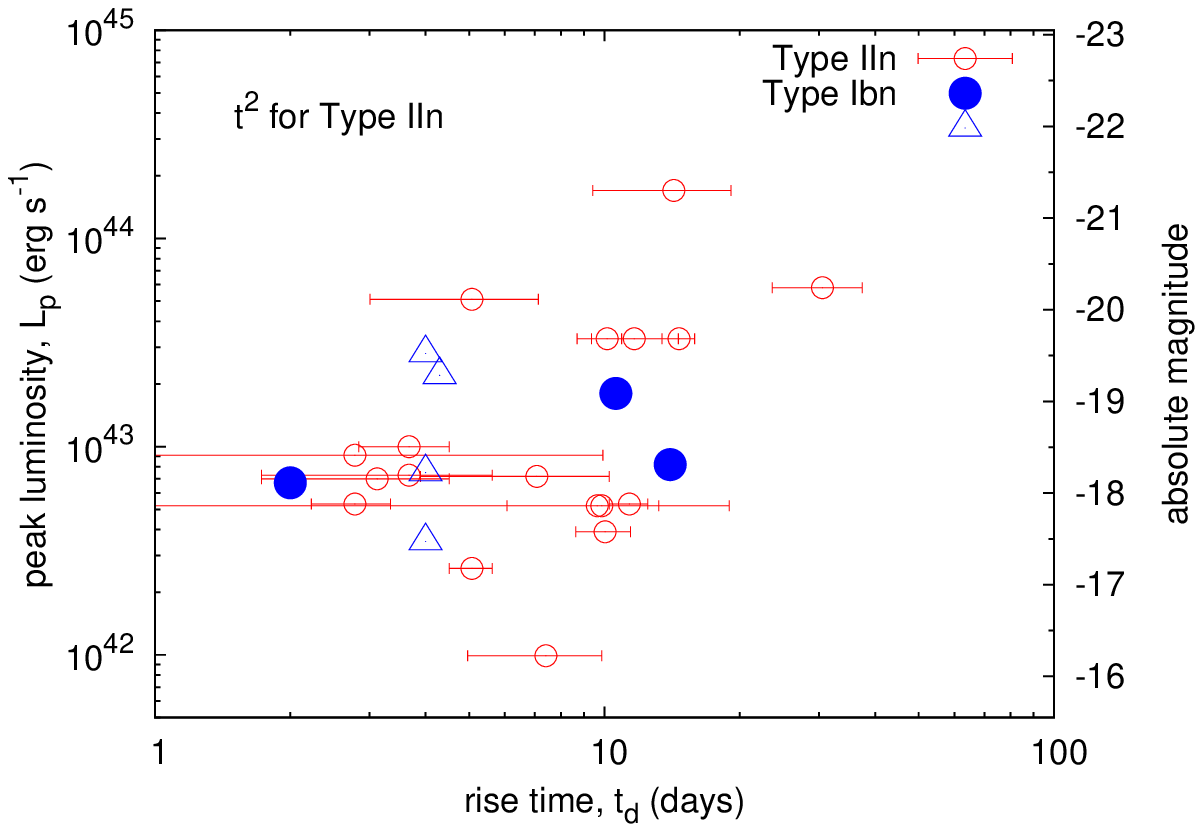} 
  \includegraphics[width=\columnwidth]{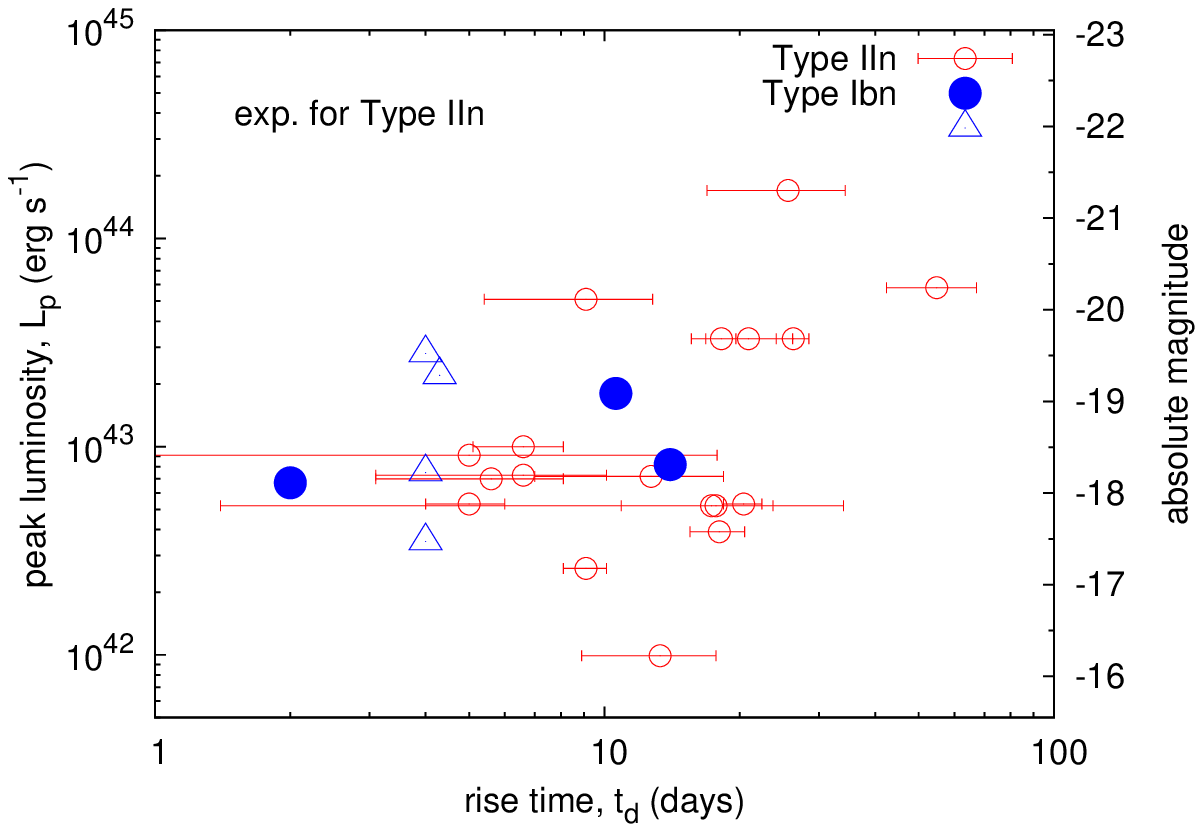}   
 \end{center}
\caption{
Rise time and peak luminosity of SNe~Ibn (Table~\ref{table:table}) and IIn. The two panels show the same data for SNe~Ibn. The left panel assume $t^{2}$ LC increase to estimate SN~IIn rise times, while the right panel assumes an exponential increase \citep{ofek2014}. SNe~Ibn in which only an upper limit for their rise time is available are plotted with open triangles and the others with filled circles.
}
\label{fig:projectb}
\end{figure*}

\begin{table*}
\begin{center}
\caption{List of SNe Ibn in Fig.~\ref{fig:projectb}.}
\begin{tabular}{lccl}
\tableline\tableline
SN name & rise time & peak luminosity & reference \\
        &   days    & $10^{42}~\mathrm{erg~s^{-1}}$ & \\
\tableline
SN 2010al   & 14   & 8.2 & \citet{pastorello2015} \\
SN 2014av   & 10.6 & 1.8 & \citet{pastorello2015d} \\
LSQ12btw    & $<4$ & 7.5 & \citet{pastorello2015b} \\
LSQ13ccw    & $<4$ & 3.5 & \citet{pastorello2015b} \\
ASASSN-15ed & $<4.3$ & 2.2 &\citet{pastorello2015e} \\
iPTF13beo   &  2   & 6.7 & \citet{gorbikov2014} \\
SN 1999cq   & $<4$ & 28 & \citet{matheson2000} \\
\tableline
\end{tabular}
\label{table:table}
\end{center}
\end{table*}

\section{Rise time and peak luminosity}\label{sec:prepeak}
We further investigate SN~Ibn properties by using their LC rise time and peak luminosity. We have previously shown that CSM and SN ejecta properties of interacting SNe are independently constrained when their rise time and peak luminosity are given, and we have estimated SN~IIn properties based on them \citep{moriya2014b}. We use the same formalism to constrain CSM and explosion properties of SNe~Ibn.

\subsection{Light-curve sample}
We take the following SNe~Ibn with a good constraint on rise time and peak bolometric luminosity. SNe~Ibn which appear in this section are summarized in Table~\ref{table:table}. SN~2010al is observed to have the rise time of 14~days and the peak luminosity of $8.2\times 10^{42}~\mathrm{erg~s^{-1}}$ \citep{pastorello2015}. SN~2014av has a rise time of 10.6~days and peak luminosity of $1.8\times 10^{42}~\mathrm{erg~s^{-1}}$ \citep{pastorello2015d}. The following SNe~Ibn have a strong constraint on their rise time and peak luminosity: LSQ12btw (rise time of less than 4~days and peak luminosity of $7.5\times 10^{42}~\mathrm{erg~s^{-1}}$, \citealt{pastorello2015b}), LSQ13ccw (rise time of less than 4~days and peak luminosity of $3.5\times 10^{42}~\mathrm{erg~s^{-1}}$, \citealt{pastorello2015b}), and ASASSN-15ed (rise time of less than 4.3~days and peak luminosity of $2.2\times 10^{42}~\mathrm{erg~s^{-1}}$, \citealt{pastorello2015e}). We note that dust formation in SNe~Ibn does not affect the peak luminosity as dusts are formed at much later phases \citep[e.g.,][]{nozawa2008}.

In addition to these SNe~Ibn in which rise time and peak bolometric luminosity are both well constrained, we add SNe~Ibn iPTF13beo \citep{gorbikov2014} and SN~1999cq \citep{matheson2000} in our sample. iPTF13beo shows a double-peaked LC, and the first peak is observed at 2~days after the explosion. This first peak is presumed to be caused by the interaction between SN ejecta and dense CSM \citep{gorbikov2014}. Only the $R$-band LC is available in the first peak. We estimate its peak bolometric luminosity as $6.7\times 10^{42}~\mathrm{erg~s^{-1}}$ by taking its peak $R$-band peak magnitude of $-18.36$~mag and assuming no bolometric correction. SN~1999cq is the first observed SN~Ibn. Its rise time is constrained to be within 4~days with the peak $R$-band magnitude of $-19.9$~mag. Assuming no bolometric correction, we obtain the peak luminosity of $2.8\times 10^{43}~\mathrm{erg~s^{-1}}$. In the following figures, we show SNe~Ibn with the upper limit on rise time with open triangles and the others with filled circles.

We use SN~IIn rise time and peak luminosity compiled by \citet{ofek2014} as in our previous study to compare SNe~Ibn with SNe~IIn. Figure~\ref{fig:projectb} shows the data. \citet{ofek2014} fit rising LCs of SNe~IIn with exponential function and provided characteristic rising timescales. They also provide a rise time assuming that rising LCs follow $t^{2}$, where $t$ is time after explosion. \citet{moriya2014b} use the latter timescale to apply their analytic formula. The two timescales provide the same conclusions, but we show the two rise times in Fig.~\ref{fig:projectb} for completeness. We only use the case of $t^{2}$ in the following discussion.

\subsection{Formalism}
We assume that shock breakout occurs in dense CSM in SNe~Ibn \citep[e.g.,][]{ofek2010,chevalier2011,gorbikov2014}. The shock breakout likely occurs within dense CSM in SNe~Ibn, because their progenitors are presumed to be compact Wolf-Rayet stars with little or no hydrogen. We discuss the validity of this assumption after we estimate the CSM density in SNe~Ibn. We assume that the outer and inner density structure of homologously-expanding SN ejecta is proportional to $r^{-n}$ and $r^{-\delta}$, respectively. Under these assumptions, we can constrain the SN ejecta and the CSM density properties independently as \citep{moriya2014b}\footnote{Eq.~(11) in \citet{moriya2014b} has a typography in the exponent of \Eej.}
\begin{equation}
\Mej^{-\frac{(4n-5)(n-5)}{2n(n-2)}}\Eej^{\frac{(4n-5)(n-3)}{2n(n-2)}}=C_3^{-1}\epsilon^{-1}\kappa^{\frac{(n-5)(n-1)}{n(n-2)}}L_pt_d^{-\frac{n^2-10n+10}{n(n-2)}},
\label{eq:snprop}
\end{equation}
\begin{equation}
D=C_2^{-\frac{n-2}{n}}C_3^{-\frac{n-2}{4n-5}}\epsilon^{-\frac{n-2}{4n-5}}\kappa^{-\frac{3(n-1)}{4n-5}}L_p^{\frac{n-2}{4n-5}}t_d^{\frac{3(n-1)}{4n-5}},
\label{eq:csmprop}
\end{equation}
where \Mej\ is SN ejecta mass, \Eej\ is SN ejecta energy, $\epsilon$ is a conversion efficiency from kinetic energy to radiation, $\kappa$ is opacity in CSM, $L_p$ is peak luminosity, $t_d$ is rise time, and $C_2$ and $C_3$ are constants which can be found in \citet{moriya2014b}. CSM density $\rho_\mathrm{CSM}$ is assumed to follow $\rho_\mathrm{CSM}=Dr^{-2}$. For the case of $n=10$, we obtain
\begin{equation}
\Mej^{-1.09}\Eej^{1.53}=C_3^{-1}\epsilon^{-1}\kappa^{0.56}L_pt_d^{-0.13},
\label{eq:snprop10}
\end{equation}
\begin{equation}
D=C_2^{-0.8}C_3^{-0.23}\epsilon^{-0.23}\kappa^{-0.77}L_p^{0.23}t_d^{0.77}.
\label{eq:csmprop10}
\end{equation}

\subsection{Results}
\begin{figure}
 \begin{center}
  \includegraphics[width=\columnwidth]{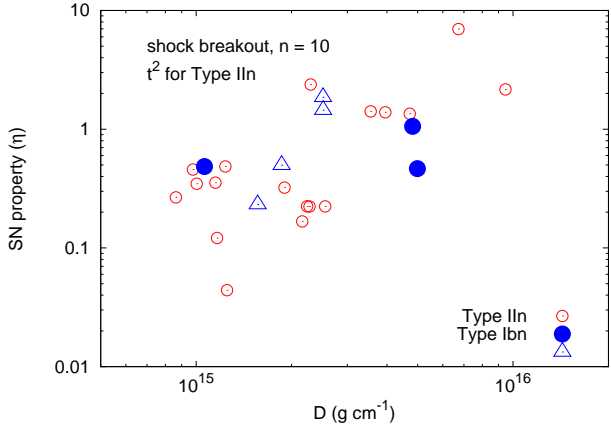}  
 \end{center}
\caption{
CSM density and SN properties of SNe~Ibn and IIn estimated by their rise time and peak luminosity.
}\label{fig:density}
\end{figure}

Figure~\ref{fig:density} shows CSM density and SN property estimated by substituting the rise time and peak luminosity in Fig.~\ref{fig:projectb} to Eqs.~(\ref{eq:snprop10}) and (\ref{eq:csmprop10}). We investigate the case of $n=10$ and $\delta=1$ in the rest of this paper, but our conclusions are not affected even if $n$ and $\delta$ are changed in a reasonable range \citep[e.g.,][]{matzner1999}. The SN property $\eta$ in Fig.~\ref{fig:density} is defined as 
\begin{equation}
\eta\equiv \left(\frac{\Mej}{5~\Msun}\right)^{-\frac{(4n-5)(n-5)}{2n(n-2)}}
\left(\frac{\Eej}{1.5\times 10^{51}~\mathrm{erg}}\right)^{\frac{(4n-5)(n-3)}{2n(n-2)}},
\label{eq:eta}
\end{equation}
or 
\begin{equation}
\eta= \left(\frac{\Mej}{5~\Msun}\right)^{-1.09}
\left(\frac{\Eej}{1.5\times 10^{51}~\mathrm{erg}}\right)^{1.53},
\label{eq:eta2}
\end{equation}
in the case of $n=10$.
We assume that $\kappa=0.2~\mathrm{cm^2~g^{-1}}$ in SNe~Ibn and $\kappa=0.34~\mathrm{cm^2~g^{-1}}$ in SNe~IIn. We also assume $\epsilon=0.3$ as in our previous study \citep{moriya2014b}.
$\epsilon$ is typically taken between 0.1 and 0.5 in the literature, and we take an average value \citep[e.g.,][]{fransson2014}.

\subsubsection{CSM density and progenitor mass loss}
We find that $D\sim 10^{15}-10^{16}~\mathrm{g~cm^{-1}}$ in SNe~Ibn (Fig.~\ref{fig:density}). Our density estimate is consistent with previously estimated CSM density in SNe~Ibn \citep[e.g.,][]{chugai2009,immler2008}. Here, we first check the validity of our assumption that shock breakout occurs in CSM in SNe~Ibn. If we assume a typical radius of Wolf-Rayet stars, $10^{11}~\mathrm{cm}$, we obtain the CSM optical depth of $\sim 10^3$ for $D\sim 10^{15}~\mathrm{g~cm^{-1}}$. This is large enough for shock breakout to occur in CSM. SNe~IIn do not necessarily experience the shock breakout in CSM because of their possible larger progenitor radii \citep{moriya2014b}, but we only show the estimated properties assuming that the shock breakout occurs in CSM in SNe~IIn here. Similar properties are obtained even if we assume that the shock breakout does not occur in SNe~IIn \citep{moriya2014b}.

An interesting result found in Fig.~\ref{fig:density} is that CSM density is similar in SN~Ibn and IIn progenitors. The typical velocities of narrow spectral components in SNe~IIn and SNe~Ibn are $\sim 100~\kmps$ \citep[e.g.,][]{taddia2013} and $\sim 1000~\kmps$ (see \citealt{pastorello2015d} for a summary), respectively. This is likely because the progenitors of SNe~IIn are presumably LBVs or red supergiants which have larger radii than hydrogen-deficit SN~Ibn progenitors (Wolf-Rayet stars). If we estimate the mass-loss rates of their progenitors, i.e., $\Mdot=D/4\pi\vw$, by assuming a typical wind velocity of $100~\mathrm{km~s^{-1}}$ for SNe~IIn and $1000~\mathrm{km~s^{-1}}$ for SNe~Ibn, the mass-loss rates of SN~Ibn progenitors become an order of magnitude higher than SN~IIn progenitors because of the higher wind velocity. Thus, a similar CSM density may indicate that the mass-loss rates of the progenitors of SNe~Ibn may be generally higher than those of SNe~IIn ($\sim 10^{-3}~\mathrm{\Msun~\mathrm{yr^{-1}}}$) by roughly a factor of $10$, i.e., $\sim 10^{-2}~\mathrm{\Msun~\mathrm{yr^{-1}}}$. However, some SNe~Ibn show narrow velocities similar to SNe~IIn and some SNe~Ibn also show hydrogen features. Thus, the difference in the mass-loss rates in SN~Ibn and IIn progenitors may not be as significant as a factor of 10, but they are likely generally higher than in SNe~IIn.

One way to interpret the possible difference in the mass-loss rates is that $D$ may need to be larger than $\sim 10^{15}~\mathrm{g~cm^{-1}}$ for SNe to show narrow lines. Thus, even if a Wolf-Rayet star has a mass-loss rate similar to those of SNe~IIn ($\sim 10^{-3}~\mathrm{\Msun~\mathrm{yr^{-1}}}$), its CSM density may not be high enough to make them classified as SNe~Ibn because of its larger wind velocity. Thus, we may only observe stripped-envelope SNe as SNe~Ibn when they have larger mass-loss rates than SNe~IIn so that they have similar CSM density. If this is the case, it implies that SN~Ib progenitors with very large mass-loss rates of $\sim 10^{-3}~\mathrm{\Msun~\mathrm{yr^{-1}}}$ may not be observed as SNe~Ibn but SNe~Ib. Then, there may exist SNe~Ib with the progenitor mass-loss rates as high as $\sim 10^{-3}~\mathrm{\Msun~\mathrm{yr^{-1}}}$ which do not show the narrow features. Thus, the fraction of Wolf-Rayet SN progenitors having large mass-loss rates may be higher than that estimated by the SN~Ibn fraction. Radio observations of SNe~Ib can eventually unveil the hidden population of Wolf-Rayet SN progenitors with large mass-loss rates, for example.
Radio observations of SNe~Ib are steadily increasing \citep[e.g.,][]{berger2003,soderberg2010} and some SN~Ib progenitors are actually found to have large mass-loss rates \citep[e.g.,][]{wellons2012}.

The difference in the rates and periods of the mass loss in SN~Ibn and SN~IIn progenitors indicates the different masses in dense CSM in the two kinds of SNe. Mass loss for more than $100~\mathrm{years}$ with $\sim 10^{-3}~\mathrm{\Msun~\mathrm{yr^{-1}}}$ in SN~IIn progenitors results in the dense CSM mass of at least $\sim 0.1~\Msun$, while mass loss for $\sim 1~\mathrm{year}$ with $\sim 10^{-2}~\mathrm{\Msun~\mathrm{yr^{-1}}}$ in SN~Ibn progenitors indicates the dense CSM mass of $\sim 0.01~\Msun$.

\begin{figure}
 \begin{center}
  \includegraphics[width=\columnwidth]{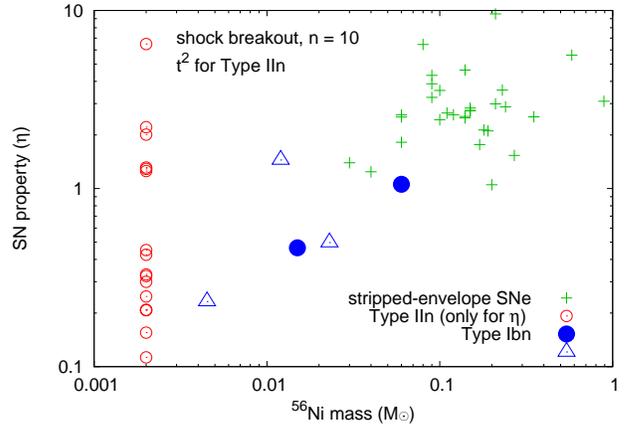}  
 \end{center}
\caption{
Estimated \Ni\ masses and explosion properties. The explosion property $\eta$ is defined in Eq.~(\ref{eq:eta}) (see also Eq.~\ref{eq:eta2}). A larger $\eta$ means a larger \Eej\ and/or a smaller \Mej. Stripped-envelope SN properties are based on \citet{cano2013}. SN properties of all SNe~IIn are shown at an arbitrary \Ni\ mass of 0.002~\Msun\ for comparison.
}\label{fig:nisn}
\end{figure}

\subsubsection{Explosion properties}\label{sec:expprop}
We further investigate explosion properties of SNe~Ibn. Figure~\ref{fig:nisn} shows the estimated \Ni\ mass and SN property $\eta$ for stripped-envelope SNe and SNe~Ibn. For stripped-envelope SNe, we take \Mej, \Eej, and \Ni\ masses estimated by \citet{cano2013} by fitting their bolometric LCs to botain the SN properties (see also \citealt{drout2011,taddia2015,lyman2014}). \Ni\ masses of SNe~Ibn are the maximum possible amount which is estimated by comparing the last bolometric luminosity available in Fig.~\ref{fig:lightcurve}a and the energy input from the nuclear decay. We also show estimated SN property $(\eta)$ of SNe~IIn for comparison. Because \Ni\ masses in SNe~IIn are hard to estimate as they are powered by the interaction for a long time, we show their $\eta$ at an arbitrary \Ni\ mass of 0.002~\Msun\ in Fig.~\ref{fig:nisn}. This \Ni\ mass does not have any meaning.

We find that \Ni\ masses constrained in SNe~Ibn are typically smaller than those of stripped-envelope SNe (see also Fig.~
\ref{fig:lightcurve}). In addition, the SN~Ibn explosion property parameters $(\eta)$ are typically smaller than those of stripped-envelope SNe. Small \Ni\ production and small $\eta$ (Eq.~\ref{eq:eta}) suggest that the explosion energy of SNe~Ibn may be generally smaller than that of stripped-envelope SNe.

As we discuss in the previous section, the dust formation in SNe~Ibn can be significant. However, small late-phase luminosity, and thus small possible \Ni\ mass, is found even SNe~Ibn with infrared observations (Fig.~\ref{fig:lightcurve}). Another concern in our way of estimating \Ni\ masses in SNe~Ibn is in the gamma-ray optical depth. If it is already below $\sim 1$ at the moment we compare the luminosity with the nuclear input energy, we will underestimate \Ni\ masses. The gamma-ray optical depth $\tau_\gamma$ at time $t$ after explosion is approximated as
\begin{equation}
\tau_\gamma \simeq \kappa_\gamma \rho_\mathrm{ej} R_\mathrm{ej}
\sim 10 \left(\frac{\eta}{0.5}\right)^{-2}\left(\frac{\Eej}{10^{51}~\mathrm{erg}}\right)^{2}\left(\frac{t}{70~\mathrm{days}}\right)^{-2},\label{eq:gammaoptdpt}
\end{equation}
where $\rho_\mathrm{ej}$ and $R_\mathrm{ej}$ are the mean ejecta density and radius, respectively. The gamma-ray opacity $\kappa_\gamma$ is set as $0.027~\mathrm{cm^2~g^{-1}}$ \citep{axelrod1980}. We assume $\eta\propto \Mej^{-1}\Eej^{1.5}$ in Eq.~(\ref{eq:gammaoptdpt}) (see Eq.~\ref{eq:eta2}). The optical depth is scaled with $\eta=0.5$, which is a mean value estimated from the rise time and peak luminosity of SNe~Ibn (Fig.~\ref{fig:nisn}). At a typical time we use for the \Ni\ mass estimate ($t\simeq 50-80~\mathrm{days}$), the optical depth is still well above one. However, we may underestimate \Ni\ masses if the explosion energy is significantly lower than $10^{51}~\mathrm{erg}$. We note that \citet{tominaga2008} present \Ni-powered bolometric LC models for SN~Ibn 2006jc by assuming very large $\Eej/\Mej$ so that the gamma-ray optical depth becomes small. However, the required $\Eej/\Mej$ results in very large ejecta velocities of more than $\sim 10000~\kmps$ which are not observed in SNe~Ibn \citep[e.g.,][]{pastorello2015d}.

The small \Ni\ mass and small explosion energy in SNe~Ibn can be interpreted in several ways. One possibility is that SNe~Ibn may suffer from large fallback and eject only small amount of \Ni\ \citep[e.g.,][]{moriya2010}. If SN~Ibn progenitors are very massive stars, it is hard to explode them and we expect small explosion energy with significant fallback which results in a small \Ni\ ejection. The fallback resulting in a small amount of \Ni\ ejection may be responsible for gamma-ray bursts (GRBs) without accompanied SNe \citep[e.g., GRB060614,][]{gehrels2006,fynbo2006,dellavalle2006,gal-yam2006,tominaga2007}. Because only little \Ni\ is ejected in this kind of SNe, we may easily observe them when this kind of progenitors experience large mass loss shortly before their explosions so that they can be bright by the interaction. Even if explosion energy is small because of the fallback, the interaction can efficiently convert the kinetic energy of ejecta to radiation with the efficiency reaching 50\% or even more depending on the mass ratio between ejecta and CSM \citep[e.g.,][]{moriya2013b} and SNe~Ibn can still become bright. We also note that if \Eej\ is significantly smaller than $10^{51}~\mathrm{erg}$, the gamma-ray optical depth becomes less than 1 earlier and we may underestimate \Ni\ masses. In this case, SNe~Ibn need to produce similar amount of \Ni\ to stripped-envelope SNe with significantly smaller explosion energies and this is unlikely \citep[e.g.,][]{umeda2008}.

Another possibility is that some SNe~Ibn may not even be related to the terminal explosions of massive stars. Some \Ni\ masses shown in Fig.~\ref{fig:nisn} are upper limits and some SNe~Ibn are consistent with no production of \Ni. Thus, we cannot rule out the possibility that some SNe~Ibn may be impostors as is often found in SNe~IIn \citep[e.g.,][]{smith2011,kochanek2012}.
Pulsational pair-instability which leads to non-terminal explosive events \citep[e.g.,][]{woosley2007,chatzopoulos2012,yoshida2016} can be related in some cases, for example.

\section{Summary and conclusions}\label{sec:summary}
We have investigated general properties of CSM and SN ejecta in SNe~Ibn by analyzing their bolometric LCs. The luminosity contrast between the LC peak and late phases in SNe~Ibn is generally much larger than that in \Ni-powered stripped-envelope SNe (Fig.~\ref{fig:lightcurve}). In other words, the \Ni\ mass required to power the peak luminosity is much larger than that inferred from the late-phase luminosity. We find this large contrast even in SNe~Ibn in which the infrared luminosity contribution is taken into account in their bolometric LCs, and the dust production is not likely the major cause of the post-peak rapid decline in many SNe~Ibn. Thus, many SNe~Ibn are not likely powered by the \Ni\ heating. Their peak luminosity is likely powered by the interaction between dense CSM and SN ejecta, as is naturally presumed from their spectral type.

When we compare LC decline rates of SNe~Ibn and SNe~IIn, we find that SNe~Ibn have much faster post-peak LC decline rates than SNe~IIn. Thus, even if both SNe~Ibn and SNe~IIn are powered by the interaction between SN ejecta and dense CSM, the interaction in SNe~Ibn is not likely sustained longer than SNe~IIn and this is why the LC decline is faster in SNe~Ibn. This indicates that SNe~Ibn have dense CSM which are less extended than those of SNe~IIn. Thus, the large mass loss making dense CSM is likely to occur in much shorter timescale close to the explosion in SNe~Ibn compared to long sustained mass loss in SNe~IIn.

The rise times and peak luminosity of SNe~Ibn and SNe~IIn are similar (Fig.~\ref{fig:projectb}). This implies that the CSM density of the two kinds of SNe is similar. Because their narrow velocity components indicating their progenitor wind velocities are often larger in SNe~Ibn, the similar CSM density may indicate that the mass-loss rates of SN~Ibn progenitors may often be larger than SNe~IIn.

The rise times and peak luminosity of SNe~Ibn also indicate that the explosion energy of SNe~Ibn is generally smaller than stripped-envelope SNe. In addition, their LCs indicate a much smaller amount of \Ni\ ejection in SNe~Ibn. Thus, SNe~Ibn may be suffering from large fallback which results in a small \Ni\ ejection. The efficient conversion from kinetic energy to radiation by the CSM interaction can make SNe~Ibn bright even if explosion energy is small because of fallback. Some SNe~Ibn may not even be a terminal explosion of massive stars and just an eruptive event with the interaction and without \Ni.

\acknowledgments{
TJM thanks the SN groups in Stockholm University and Padua Observatory for fruitful discussion.
TJM is supported by Japan Society for the Promotion of Science Postdoctoral Fellowships for Research Abroad (26\textperiodcentered 51).
The work by KM is supported by Japan Society for the Promotion of Science
(JSPS) KAKENHI Grant (26800100), and by World Premier International Research
Center Initiative (WPI Initiative), MEXT, Japan.
}

\end{document}